\documentclass[runningheads]{llncs}
\usepackage[T1]{fontenc}
\usepackage{graphicx}
\usepackage{amsmath}
\usepackage{amssymb}
\usepackage{amsfonts}
\usepackage{ulem,cancel}
\usepackage{cleveref}

\DeclareMathOperator{\sign}{sign}
\DeclareMathOperator{\clamp}{clamp}

\usepackage[bibliography=common]{apxproof}

\begin{document}
\title{Modular population protocols}
%
%
\author{Michael Raskin
\orcidID{0000-0002-6660-5673}
}
%
\authorrunning{M.Raskin}
%
\institute{LaBRI, University of Bordeaux, CNRS UMR 5800
\email{mraskin@u-bordeaux.fr}
}
\maketitle              
\begin{abstract}
Population protocols are a model of distributed computation intended for the study of networks of independent computing agents with dynamic communication structure.  Each agent has a finite number of states, and communication opportunities occur nondeterministically, allowing the agents involved to change their states based on each other's states. Population protocols are often studied in terms of reaching a consensus on whether the input configuration satisfied some predicate.

A desirable property of a computation model is modularity, the ability to combine existing simpler computations in a straightforward way.  In the present paper we present a more general notion of functionality implemented by a population protocol in terms of multisets of inputs and outputs.  This notion allows to design multiphase protocols as combinations of independently defined phases.  The additional generality also increases the range of behaviours that can be captured in applications (e.g. maintaining the role distribution in a fleet of servers).

We show that composition of protocols can be performed in a uniform mechanical way, and that the expressive power is essentially semilinear, similar to the predicate expressive power in the original population protocol setting.

\keywords{Population protocols \and Protocol verification \and Modularity.}
\end{abstract}

\section{Introduction}

\subsection{General context of population protocols}

Population protocols have been
introduced in \cite{conf/podc/AngluinADFP04,journals/dc/AngluinADFP06}
as a restricted yet useful subclass of general distributed protocols.
Each agent in a population protocol has a fixed amount of local storage,
and an execution consists of 
selecting pairs of agents and letting them update their states
based on an interaction.
The choice of pairs is assumed to be performed by an adversary 
subject to a fairness condition.
The fairness condition ensures
that the adversary must allow the protocol to progress.

Typically, population protocols are studied from the point of view
of recognising some properties of an input configuration.
In this context population protocols and their subclasses have been studied,
for example,
from the point of view of expressive power \cite{journals/dc/AngluinAER07},
verification complexity
\cite{conf/podc/BlondinEJM17,esparza2019parameterized,DBLP:journals/dc/EsparzaJRW21},
time to convergence \cite{conf/wdag/AngluinAE06,conf/wdag/DotyS15},
necessary state count \cite{blondin2019succinct},
etc. 

The original target application
of population protocols and related models
is modelling networks of restricted sensors,
starting from
the original paper \cite{conf/podc/AngluinADFP04} on population protocols.
Of course, in the modern applications the cheapest microcontrollers
typically have thousands of bits of volatile memory
permitting the use of simpler and faster algorithms for recognising 
properties of an input configuration.
So on the one hand, the original motivation for the restrictions in the 
population protocol model seems to have less relevance.
On the other hand, verifying distributed systems
benefits from access to a variety of restricted models
with a wide range of trade-offs between
the expressive power and verification complexity,
as most problems are undecidable
in the unrestricted case.
Complex, unrestricted, and impossible to verify distributed deployments
lead to undesirable and hard to predict and sometimes even diagnose situations 
such as so called gray failures \cite{Huang_2017} and similar.

From the theoretical point of view,
population protocols provide a model
of distributed computing with some of verification problems
still decidable\cite{esparza_et_al:LIPIcs:2016:6862}
although non-elementary\cite{CLLLM19}.
This property puts them near the edge of decidability of verification,
which is interesting on its own.

\subsection{Composition}

Unfortunately, as the standard approach considers protocols
that compute predicates, composing protocols directly is limited 
to boolean combinations.
Whenever a multiphase protocol needs to be constructed,
the interaction between phases needs to be described and proven
in an ad hoc way.
We find it desirable to have better modularity in protocol design.

Naturally, for the composition to be useful, we need the protocols
to achieve more complex output distributions than consensus.
Moreover, some constructions in the literature \cite{DBLP:journals/jcss/CzernerGHE24,DBLP:conf/opodis/GasieniecHMSS16}
have more or less sub-protocols executed in parallel with different agents
participating in different sub-protocols based on the output of yet another 
«previous» sub-protocol.
From the other point of view,
desirable behaviours of distributed systems
go beyond
consensus about a single property of initial configuration.
For example, one might want to maintain task allocation
across a fleet of servers.
Task allocation usually depends on the server models,
and servers are sometimes taken out of service for maintenance.
From this point of view,
we also find it interesting to study a more general notion of expressive power.

\subsection{Related work}

Many approaches to composition of distributed protocols 
depend on 
a fixed communication structure \cite{Austin2018CompositionalAT},
relatively rich local capabilities with infinite state space \cite{Viroli2017EngineeringRC},
or specific limits on how often each agents must be scheduled \cite{DelporteGallet2007RobustSL}.
As there is no natural translation of such approaches to the model of population protocols, 
and the possibilities are less restricted in the richer models, 
we do not aim to classify such approaches.

Many composition methods rely on the notion of self-stabilisation 
introduced by Dijkstra \cite{Dijkstra1974SelfstabilizingSI}, informally
meaning that the agents will reach a correct (collective) configuration
no matter what are their initial local states.
Self-stabilising protocols compose naturally. Indeed, whatever effects
on the consumer protocol happen because of the changes in the output
of the producer protocol, the end results can be treated as the initial 
configuration before the self-stabilisation in the consumer protocol starts.
This has been long used to design
composition formalisms \cite{92911,DBLP:journals/dc/DolevIM93}.
Self-stabilisation has been applied to
population protocols \cite{Angluin2005SelfstabilizingPP}.
There, a general definition of behaviours as sets of series of 
output configurations is used, and a subclass behaviours
with eventually unchanging outputs is defined.
However, self-stabilisation is a very restrictive requirement for
a population protocol where each agent may talk to each other
(i.e. the communication graph is a clique). 
It is shown \cite{Angluin2005SelfstabilizingPP} 
that one cannot do leader election in a self-stabilising way,
and the same proof applies to reaching the consensus whether 
there is currently a leader.

Some papers on self-stabilisation use lemmas about composing
completely generic behaviours \cite{DBLP:journals/dc/DolevIM93}, 
but this is done in a low-level way.

The limitations of self-stabilisation for population protocols
are not shared by 
the protocols with stabilising inputs \cite{Angluin2005StablyCP}.
The idea is that each agent gets an input that can change
finitely many times.
The authors mention that this approach is useful for composition,
but only study reaching consensus on the value of some predicates.
Moreover, only a conference version of that article exists with 
proofs deferred to a full version that has never been finished,
so for some statements it is unclear whether they should be 
considered proven.

One of the works on majority \cite{DBLP:conf/opodis/GasieniecHMSS16}
explicitly describes the most complicated protocol via a sequential 
composition of protocols with eventually-stabilising inputs, while
solving a problem that is impossible for general self-stabilising 
protocols. 
However, the proofs are specific to the protocols used and no 
general statement (nor a general definition) is given.
In a sense, our approach is a broad and reusable generalisation of 
the approach there.

\subsection{Contribution}

In the present paper we introduce an alternative 
general notion of a protocol
implementing an input-output specification.
We consider each agent to have an input,
and as usual there is an output function.
An input-output  specification 
is a relation between multisets of inputs and multisets of outputs,
plus a compatibility relation between individual inputs and outputs.
A protocol implements the specification, if in each fair execution
from an input configuration
there is a step after which the specification is always satisfied,
both in terms of the multisets of inputs and outputs, 
and in terms of each agent's output being compatible with the input.
The standard notion of computing predicates by stable consensus
corresponds to a natural class of specifications
where output is a single bit, only consensus output configurations
are acceptable,
and for each input configuration only one of the two consensus options
is acceptable.

Additionally, we define computations with shutdown requests.
We allow the scheduler to add new agents, or to tell existing agents
to leave the computation.
Each leaving agent can take some time to hand over the information,
but must eventually terminate (in a fair execution).
Once there is no turnover, an input-output specification
must be eventually stably satisfied.

We show a natural completely mechanical way of combining
two population protocols with shutdown requests implementing two relations
to obtain a protocol with shutdown requests implementing 
the composition of these relations.

In terms of handling the inputs,
we add the possibility of shutdown request to 
the approach of stabilising inputs.
In terms of the specifications implemented,
we study what multisets of outputs can get obtained from what
multisets of inputs, while previous work for stabilising
inputs only considered computing predicates.

The rest of the present paper is organised as follows.
We start with presenting the standard definitions for population
protocols, then in the next section we define our extension to 
the model to speak about specifications beyond predicate evaluation
and provide some examples.
In the section after that we define and construct sequential composition of protocols.
We then proceed to characterise the expressive power of population protocols
under our definition.
The paper ends with a small conclusion with some future directions outlined.

 For better alignment with the conference version, 
 the proofs missing in the main text 
 have been placed in the appendix.

\section{Basic definitions}

In this section we recall the standard definitions 
and facts
related to population protocols.
We use the definition where agents have identities for the purposes
of analysis but cannot distinguish each other.
This will also be relevant in the generalisations.
First we define the population protocols.
We start by describing what information we need to specify a protocol.

\begin{definition}
        A
        \emph{population protocol}
        is defined by a finite set of 
        \emph{states}~$Q$
        and a \emph{step relation}~$Step\subset{}Q^2\times{}Q^2$.
        When there is no ambiguity about the protocol,
        we abbreviate $((q_1,q_2),(q'_1,q'_2))\in{}Step$
        as $(q_1,q_2)\mapsto(q'_1,q'_2)$
        and call the quadruple
        $(q_1,q_2)\mapsto(q'_1,q'_2)$ a \emph{transition}.
\end{definition}

We use the following notation to work with
function on agents.

\begin{definition}
For a function $f$ let $\mathrm{Dom}(f)$ denote the domain of the function.

        For a function $f$ and $x\notin \mathrm{Dom}(f)$ let $f\cup\{x\mapsto y\}$
        denote the function $g$ defined on $\mathrm{Dom}(f)\cup \{x\}$ such that
        $g\mid_{\mathrm{Dom}(f)}=f$ and $g(x)=y$.
        For $u\in \mathrm{Dom}(f)$ let $f[u\mapsto v]$ denote
        the function $h$ defined on $\mathrm{Dom}(f)$ such that
        $h\mid_{\mathrm{Dom}(f)\setminus\{u\}}=f\mid_{Dom(f)\setminus\{u\}}$ and $h(u)=v$.
        For symmetry, if $w=f(u)$
        let $f\setminus\{u\mapsto w\}$ denote
        restriction $f\mid_{\mathrm{Dom}(f)\setminus\{u\}}$.

        Use of this notation implies an assertion of correctness,
        i.e. $x\notin \mathrm{Dom}(f)$, $u\in \mathrm{Dom}(f)$, and $w=f(u)$.
\end{definition}

\begin{definition}
        A \emph{configuration}
        of a population protocol is a set $A$ of agent identites,
        and a state function $C:A\to Q$, assigning states to agents.
We identify the configuration with the state function as $A=\mathrm{Dom}(C)$.
        The \emph{size} of a configuration~$C$
        is the number of agents.
        An \emph{execution} of a population protocol
        is a finite or infinite sequence~$(C_0, C_1. \ldots)$
        of configurations such that the set of agents is the
same for each $C_j$, 
and for each $j$ between $1$ and the execution length we have
some two agents $a_1, a_2$
that interact according to the rules,
i.e. $C_{j}(a_1) = q_1$, $C_j(a_2) = q_2$,
$C_{j+1}=C_j[a_1\mapsto q'_1][a_2\mapsto q'_2]$
        where $((q_1,q_2),(q'_1,q'_2))\in{}Step$.
        In other words, we let two agents with states
        $q_1$ and $q_2$ interact in some way permitted by the $Step$ relation.
\end{definition}

\begin{example}
        \label{ex:simple-protocols}
        Consider the set of states $\{q_0,q_1,q_2,q_3\}$.
        The step relation is described by
\begin{eqnarray*}
        (q_1,q_1)\mapsto(q_0,q_2),\\
        (q_2,q_1)\mapsto(q_0,q_3),\\
        (q_2,q_2)\mapsto(q_1,q_3),\\
        (q_0,q_3)\mapsto(q_3,q_3),\\
        (q_1,q_3)\mapsto(q_3,q_3),\\
        (q_2,q_3)\mapsto(q_3,q_3).
\end{eqnarray*}
        Using agent identities $1$, $2$, and $3$, 
and denoting a function from ${1,2,3}$ as a tuple,
we can consider the following execution:
$(q_1, q_1, q_1),\allowbreak (q_2, q_1, q_0),\allowbreak
(q_0,q_3,q_0),\allowbreak (q_3, q_3, q_0),\allowbreak 
(q_3,q_3,q_3)$.
        Here we use the first two steps, then twice the fourth step.
\end{example}

\begin{remark}
Note that all the configurations in an execution have the same size.
\end{remark}

We often consider executions with the steps chosen by an adversary.
However, we need to restrict adversary to ensure 
that some useful computation remains possible.
To prevent the adversary from e.g. only letting one pair of agents to interact,
we require the executions to be fair.
The fairness condition can also be described 
by comparison with random choice of steps to perform:
fairness is a combinatorial way to exclude a zero-probability set of bad executions.

\begin{definition}
        Consider a population protocol $(Q,Step)$.

        A configuration~$C'$ is \emph{reachable} from configuration~$C$ 
        if{}f there is a finite execution
        with the initial configuration $C$
        and the final configuration $C'$.

        A finite execution is \emph{fair}
        if it cannot be extended, i.e. it is not a prefix
	of any longer execution.

        An infinite execution $C_0, C_1, \ldots$ is \emph{fair}
        if for every configuration~$C'$
        either~$C'$ is not reachable from some~$C_j$
        (and all the following configurations),
        or $C'$ occurs among $C_j$ infinitely many times.
\end{definition}

\begin{example}
        The finite executions in the example~\ref{ex:simple-protocols} are fair.
\end{example}

\begin{remark}
        As long as the number of reachable configurations
        is finite, fairness implies that every set of configurations
        either becomes unreachable, or gets reached infinitely many times.
\end{remark}

The most popular notion of expressive power for 
population protocols is computing predicates,
defined in the following way.

\begin{definition}
        Consider a population protocol $(Q,Step)$
        with additionally defined
        non-empty set of \emph{input states}~$I_s\subset{}Q$,
        output alphabet $O$
        and \emph{output function} $o: Q\to{}O$.

	A state is 
        \emph{inhabited} in the configuration
	if there is at least one agent with this state.
        Considering a configuration as a function from
        agents to states, we can say that a state $q$
	is inhabited if
	$q \in \mathrm{Im}(C)$.

        A configuration $C$ is an \emph{input configuration}
	if all the inhabited states are among the input states,
	$\mathrm{Im}(C)\subset I_s$.

        A configuration $C$ is a $b$-\emph{consensus}
        for some $b\in{}O$
        if the output function yields $b$
        for all the inhabited states,
	$\mathrm{Im}(o\circ C)=\{b\}$.
        A configuration is a \emph{stable $b$-consensus}
        if it is a $b$-consensus together 
        with all the configurations reachable from it.
        A configuration is called 
        just a \emph{consensus}
        or a \emph{stable consensus}
        if it is a $b$-consensus (respectively stable $b$-consensus)
        for some $b$.

        A protocol \emph{computes} a function
        $\varphi:\mathbb{N}^I_s\to{}O$
        if{}f 
        for each input configuration~$C$
        every fair execution with initial configuration $C$
        contains a stable $\varphi(C)$-consensus.
        We usually use the protocols computing predicates,
        which corresponds to $O=\{true,false\}$.
\end{definition}

\begin{example}
        If we define the set of input states $I_s=\{q_1\}$,
        output set $O=\{true,false\}$
        and the output function $o(q){=}(q=q_3)$,
        the protocol in the example~\ref{ex:simple-protocols}
        computes the predicate $\varphi(C)=(C(q_1)\geq{}3)$.
\end{example}

\section{Model extensions}

In this section we describe our extensions.

Informally speaking, each agent knows its input
(either a value or a shutdown request) 
as a part of the state,
but the agent cannot change it.
For convenience, all input state 
have the same «empty» internal component of the state.
An agent whose input becomes a shutdown request
and doesn't change should eventually 
change the internal component of the state to empty;
agents with a shutdown request and empty internal
component of the state do not participate in any interactions
(and can be removed from the configuration).
The scheduler can change the inputs 
and add agents in the input states, 
but only finitely many times.

A specification consists of two relations:
the main relation
on multisets of inputs and multisets of outputs;
and a secondary relation on the input language and the output language.
The first relation describes the global-level requirements,
and the second describes compatibility of inputs and outputs
(e.g. a server needs to have enough storage if it is a backup server).
For every multiset of inputs 
there must be at least one acceptable multiset of input-output pairs
(i.e. the task is never completely impossible).
A protocol implements a specification $(\varphi,\psi)$
if any finite sequence of scheduler actions
from an input configuration followed by a fair execution
without input changes or agent addition/removal
eventually has a stable output for each agent
and the multisets of inputs and outputs
satisfy the relation $\varphi$, while each agent's
input and output satisfy $\psi$.
So we could say something like «the input being amount of storage for a server,
we consider as many webservers as there are servers without sufficient storage
for a database server, as many database servers as webservers,
and the remaining servers as backup servers — and only servers 
with enough storage can be database servers or backup servers».
Then after decommissioning some servers,
deploying some new ones, and reconfiguring a few of the remaining ones,
the fleet settles into a role assignment compliant with the policy.
Formally, we also need some policy what to do when no servers have sufficient storage,
accepting something that would be a misconfiguration otherwise.

\begin{definition}
        We use $\bot$ to denote lack of value 
        (in the sense of a shutdown request or 
        in the sense of the initial empty internal state).

        A protocol is \emph{input-saving} if there is a set of input values $I$ 
        such that the set of states 
        $Q$ is the Cartesian product 
        $(I\cup\{\bot\})\times(Q_{int}\cup\{\bot\})$,
        the set of input states $I_s$ is equal to $I\times\{\bot\}$,
        the output function on $(\bot,\bot)$ is equal to $\bot$,
        and the following conditions on the step relation hold:
\begin{enumerate}
        \item the first component of the state (the \emph{input}) 
                doesn't change;
        \item if at least one agent has both components of the state
                equal to $\bot$, the states don't change at all
                (the agents don't interact while shut down).
\end{enumerate}
        We call the first component of the state the \emph{input},
        and the second component of the state
        the \emph{internal memory}.

        A \emph{reconfiguration} of an input-saving population protocol is one 
        of the following changes to the configuration:
\begin{enumerate}
        \item removing an agent in the shutdown state $(\bot,\bot)$;
        \item adding an agent in the shutdown state $(\bot,\bot)$;
        \item changing an agent's input, i.e. replacing a state $(i,q)$
                with a state $(i',q)$ that differs only in the first component.
\end{enumerate}

        An \emph{execution with reconfiguration} of an input-saving protocol
        is a finite or infinite sequence of steps and reconfigurations.
        An execution with reconfiguration is \emph{fair} if it has a finite
        number of reconfigurations, and its suffix after the last reconfiguration
        is a fair execution.

        An input-saving population protocol \emph{respects shutdown requests},
        if in each fair execution with reconfiguration
        after some time 
        either
        all agents have the input $\bot$
        or
        all the agents with the input $\bot$ are in 
        the shutdown state $(\bot,\bot)$.

        An input-saving population protocol 
        \emph{implements} a specification $(\varphi,\psi)$
        if the following conditions all hold:
\begin{enumerate}
        \item the protocol respects shutdown requests;
        \item in each fair execution with reconfiguration
                each agent's output 
                only changes a finite number of times;
        \item $\varphi$ is satisfied by the multiset of inputs and 
                the multiset of outputs after stabilisation of outputs;
        \item for each agent, $\psi$ is satisfied
                by the agent's input and output after stabilisation.
\end{enumerate}
\end{definition}

\begin{remark}
If we want an execution with reconfiguration to start
with a non-empty configuration,
we can instead use reconfiguration to add the necessary agents in the
very beginning of the execution.
\end{remark}

\begin{remark}
After the last reconfiguration in a fair execution with reconfigurations
the number of reachable configurations becomes finite, 
so by fairness any set of configurations has to become unreachable
or be reached infinitely many times.
\end{remark}

\begin{remark}
When verifying a specific protocol, $\psi$ is usually 
trivially ensured by the output function.
\end{remark}

\begin{example}
        \label{example:presence}
        Consider the following input-saving protocol.

        The input language has two elements, «Yes» and «Maybe».
        The output language has three elements, 
        «Yes» and «No», and $\bot$.
        The internal memory language has four elements
        «Me», «Yes», «No», and $\bot$.

        The output function returns $\bot$ when the input 
        is $\bot$,
        «Yes» when the input is «Yes» or the internal memory
        is «Yes»,
        and «No» otherwise.

        In a pattern form:
\begin{eqnarray}
        (\bot, *) & \to & \bot \\
         (Yes, *) & \to & Yes    \\
         (Maybe, Yes) & \to & Yes  \\
         (Maybe, Me/No/\bot) & \to & No 
\end{eqnarray}

        The step function works as follows 
        (right-hand-side $*$ keeps the value from the left-hand-side;
        same interactions are also possible in the reverse order).
\begin{eqnarray}
        (\bot, *), (Yes, *) & \to & (\bot, \bot), (Yes, Me)
        \\ (\bot, Me), (Maybe, *) & \to &  (\bot, \bot), (Maybe, Me)
        \\ (\bot, Yes/No), (*, *) & \to & (\bot, \bot), (*, *)
        \\ (Yes, *), (Yes, *) & \to & (Yes, Me), (Yes, Me)
        \\ (Yes, *), (Maybe, *) & \to & (Yes, Me), (Maybe, Yes)
        \\ (Maybe, Me), (Maybe, *) & \to & (Maybe, Me), (Maybe, No)
        \\ (Maybe, \bcancel{Me}), (Maybe, \bcancel{Me}) & \to & (Maybe, *), (Maybe, *)
\end{eqnarray}
Here $\bcancel{Me}$ means any value allowed in this position except $Me$.

Informally, internal memory «Me»
is set by an own «Yes» input and 
gets replaced with «Yes» by observing another «Yes» input 
while not having a «Yes» input;
it is also «handed over» before the shutdown;
internal memory «Yes» is set 
by observing a «Yes» input of another agent
and removed by observing a «Me» internal memory 
of an agent without «Yes» input.

We observe the following.
\begin{enumerate}
        \item The protocol respects shutdown requests.
        \item The protocol implements the specification
                «all inputs are ‘‘Maybe’’ and all outputs are ‘‘No’’,
                or there is a ‘‘Yes’’ input and all outputs are ‘‘Yes’’»
\end{enumerate}
\end{example}

The basic idea is that «Yes» in the memory appears when «Me» appears somewhere else,
eventually all «Me» are for the same input, and «Me» with a given input spread 
either «Yes» or «No» output.

\begin{toappendix}
        \begin{proof}[of \cref{example:presence}]
\begin{enumerate}
        \item Any interaction of an agent with the input $\bot$ 
                with an agent with a non-shutdown input
                leads to the agent with the input $\bot$ shutting down.
                Any fair execution will eventually need to include
                such an interaction given that there is an agent with
                $\bot$ input and an agent with a non-$\bot$ input.
        \item
                First, observe that as long as there is a 
                («Maybe»,«Yes») agent there is an agent 
                with «Me» in the internal memory.
                Indeed, any creation of a state («Maybe»,«Yes») also 
                ensures existence of 
                («Yes»,«Me»); while reducing the number of «Me»
                is only possible by replacing two of them with one.
                As the protocol cannot create a («Yes»,«Yes»)
                with a step, reconfigurations also preserve this
                condition. 

                Second, observe that as long as there is a «Yes» input
                after the last reconfiguration, interacting with this
                agent reduces the number of («Maybe», «Me»).
                Therefore eventually there will be no «Yes» inputs or
                no («Maybe», «Me»).

                Third, «Yes» inputs convert «Maybe» inputs to 
                («Maybe», «Yes»), while («Maybe», «Me») converts
                «Maybe» inputs with non-«Me» internal memory to
                («Maybe», «No»).
                Therefore either there is never a «Me» internal memory,
                or eventually everyone has output «Yes» forever,
                or eventually
                everyone has output «No» forever.
                In particular, the outputs of all agents stabilise.
\end{enumerate}
\end{proof}
\end{toappendix}

Leader election can be implemented in a similar way, 
and we provide the details in the 
 appendix.
We only promise that eventually there will be a leader if after reconfigurations there are at least two
non-shutdown agents.

\begin{toappendix}
\begin{example}
        \label{example:leader}
        Consider the input-saving protocol with input language of one symbol $\top$
        (the shutdown request $\bot$ being added to the input language),
        output language $\{Leader,Follower\}$, and the same internal memory language
        (plus $\bot$).
        The output function returns the internal memory state.
        The transitions are:
\begin{eqnarray}
        (\bot,*), (\top,Leader) & \to & (\bot,\bot),(\top,Leader) \\
        (\bot,Leader), (\top,*) & \to & (\bot,\bot),(\top,Leader) \\
        (\bot,Follower), (\top,\bcancel{Leader}) & \to & (\bot,\bot),(\top,Follower) \\
        (\bot, \bcancel{Leader}), (\bot,\bcancel{Leader}) &\to& (\bot,\bot),(\bot,\bot) \\
        (\bot, Leader), (\bot,*) &\to& (\bot,Leader),(\bot,\bot) \\
        (\top, Leader), (\top, *) &\to& (\top,Leader), (\top,Follower)\\
        (\top, Follower), (\top, \bcancel{Leader}) &\to& (\top,Follower), (\top,Follower)\\
        (\top, \bot), (\top, \bot) &\to& (\top,Leader), (\top,Follower)
\end{eqnarray}
Informally, Follower knows there is a Leader, one of two Leaders becomes a Follower on collision,
and when there is no information one of the agents opportunistically declares itself a leader.
During the shutdown, the agent makes sure to pass the Leader status to another agent.
\end{example}
        \begin{proof}[of \cref{example:leader}]
An agent with a shutdown request as the input will shutdown after any interaction 
with an agent without a shutdown request.
A Follower can only exist if there is a Leader, the number of Leaders cannot 
go to zero from positive, but the number of Leaders can only decrease
once there are no $(\top,\bot)$ agents and no reconfigurations happen.
Thus the protocol implements leader election.
\end{proof}
\end{toappendix}

\begin{example}
        \label{example:sum}
Consider the following protocol, parametrised by a positive integer $m$.
        The input language is $\{-m, -m+1, \ldots, m-1, m\}$.
        The output language is the same as the input language.
        The internal memory language is 
$\{-m, -m+1, \ldots, m-1, m\}\times\{-2m, -2m+1, \ldots, 2m-1, 2m\}$.
We call the two sub-components of the internal memory
«previous input» and «current balance».

We will call $k$-clamping of a number $x$ the number $x$ itself
        if it is at most $k$ by absolute value (i.e. $|x|\leq k$)
        and $k$ or $-k$ according to the sign of $x$ otherwise
        (i.e. $k \cdot \sign x$).
        We write $\clamp_k x=min(k,max(-k,x))$.

The output function is the $m$-clamping of the current balance.

The step function is as follows.

First, for each of the agents 
we consider the current input $i$, the previous input $p$ and 
the current balance $b$.
We try to add the difference between the previous and the current 
input to the current balance and to the previous input, but make sure
        the values stay in the permitted range.
        More precisely, we consider $b'=\clamp_{2m}(b+i-p)$,
        $p'=p+b'-b$, and update the current balance to become $b'$ 
        and the previous input to become $p'$.
        (As $p'$ is between $p$ and $i$, it cannot cause range problems).

Then we consider the two updated current balances, $b'_1\leq b'_2$
        (possibly swapping the agents to ensure that the first balance is larger).
        We replace them with $b''_1=\clamp_{2m} (b'_1+b'_2)$ 
        and $b''_2 = b'_1 + b'_2 - b''_1$.

Then if exactly one of the agents has the current input $\bot$
and the other agent has the current balance $0$, we swap the current balances.
Afterwards each agent with the current input $\bot$, the previous input $0$,
and the current balance $0$ gets the internal memory set to $\bot$ to 
        shut it down.

Observe that this protocol 
        implements the specification that either there is at most one
        agent, or the following holds.
\begin{itemize}
        \item Either there are no agents with a negative output
                or there are not agents with a positive output.
        \item The largest output by absolute value is 
                the (clamped) sum of all the current inputs
                $\clamp_{m}\sum_j i_j$. 
\end{itemize}

\end{example}

The basic idea is that the current balances cancel out when possible
and accumulate otherwise, while the unused difference between
current input and previous input 
in a stable situation
can only be in the same direction as the
sign of the global sum of inputs.

\begin{toappendix}
        \begin{proof}[of \cref{example:sum}]
        First of all, the sum of the current balances 
                over all the agents
                is always 
        equal to the sum of the saved previous
        inputs, as reconfigurations 
                cannot change either value
                and steps can only change them by the same amount.

         Observe that once the reconfigurations cease,
        the sum of absolute values of differences between
        the current input and the stored previous input 
        can only decrease.
        Thus when analysing a fair execution with reconfiguration
        we can assume that the sum of absolute values of these
        differences has achieved its minimum at some point 
        and consider the execution afterwards.

        Note that the difference between
        the current and the previous inputs is always of the same sign
        as the current balance from this point on.
        Indeed, otherwise the agent would decrease the difference 
        at the next interaction.

        Now consider the sum of the absolute values of the current balances.
        It can only grow due to updating the balance and the previous input,
        decreasing the sum of absolute values of differences between the
        current and the previous inputs. Once the latter sum has stabilised,
        the former sum can only decrease, so we can consider the execution
        once it has achieved its minimum.
        As an interaction between the agents with the balances of the opposite
        signs will decrease that sum, the agents' current balances, and thus
        outputs, have to to be all non-negative or all non-positive from this
        point on.

        Now consider the number of agents that have non-zero non-maximum 
        current balance. As by now the current balances can only shift around
        between the agents, the number of such agents can only decrease.
        Moreover, at this point 
        interaction of two such agents leaves only one such agent
        (merging the balances either creates the maximum value or 
        a zero value).
        Thus eventually there will be at most one such agent and we can consider
        the execution starting from this point.

        Observe that the number non-shut-down agents with
        the current input $\bot$ can only decrease.
        Note that the sum of all current balances is always the sum
        of the stored previous inputs,
        and by the current moment in the execution it 
        has the absolute
        value at most the sum 
        of the absolute values of current inputs 
        (we have already shown that 
        the difference can only go in one direction by this point).
        As the current input $\bot$ counts as zero, and each 
        agent can hold more as the current balance 
        than as the previous input, as long as there is a non-shut-down
        agent with a shutdown request, there is an agent with the current
        balance $0$.
        Interaction of these two agents lets the agent with the shutdown 
        request to get current balance to $0$; if the previous input 
        were non-zero, the difference could decrease
        but we are in the part of the execution where this is impossible
        thus the agent can be shut down completely, decreasing the number
        of such agents.
        Thus as long as not all agents have shut down requests, all
        the shutdown requests will be honoured.
        
        Now observe that either we have an agent with the current
        balance at least $m$ by absolute value and thus output $\pm m$,
        which has to coincide with the clamped sum of the current inputs;
        or we have 
        a single non-zero-balance agent (and then its current balance
        has to be equal to the sum of the current inputs).

        Finally, once the number of agents with neither 
        zero nor extreme internal memory stabilises at 
        zero or one, and there are no outstanding shutdown
        requests, even the internal memory states stop changing.
        A fortiori, the outputs of the agents stabilise by that point.
\end{proof}
\end{toappendix}

\begin{remark}
If we replace clamping with taking a remainder modulo $m$,
and all the sums in the specification and the proof with sums modulo $m$,
we obtain a very similar protocol for computing the sum of the inputs modulo $m$
as the only non-zero agent output.
\end{remark}

\subsection{Parallel composition}

In this section we briefly mention the simpler composition methods
for input-saving population protocols.

If we have two protocols with the same input language,
we can take the Cartesian product of their output languages
and internal memory languages, and execute the two protocols
in parallel on the same population.
For the internal memory we identify  $(\bot,\bot)$ with $\bot$.
Clearly, the product protocol will respect the shutdown requests
if the original ones did.
If the protocols implemented specifications $(\varphi_1,\psi_1)$ 
and $(\varphi_2,\psi_2)$,
the product protocol
will implement the specification
that accepts a multiset of inputs and a multiset of outputs if 
replacing outputs with their first components makes
$\varphi_1$ accept the pair of multisets 
and $\psi_1$ accept each agent's input-output pair, 
while replacing outputs with their second components
satisfies $\varphi_2$ globally and $\psi_2$ locally.

It is possible that the protocols we wanted to combine 
had different inputs, and we want to combine the outputs
in some way other than building a pair. 
For the output side, we can replace the output function
with a composition of the old output function and an output-translating
function.
For the input side, given an input-translating function, we need
to define the new transitions as the old transitions, where
inputs are the images of the actual inputs.
The specification will be composed with the translation functions
in a natural way.

\begin{example}
        We have a protocol (\cref{example:presence})
        to check whether any agent has input «Yes». 
        If we want to verify that with the input
        language $\{0,1,2\}$ some agents 
        have $0$ and some agents have $1$,
        we can ask each agent to participate in two
        independent copies of the «Yes»-checking protocol.
        One copy would interpret $0$ as «Yes» (other inputs as «Maybe»),
        and the other copy would interpret $1$ as «Yes».
        The output of naive Cartesian product protocol
        would be two «Yes»/«No» values,
        but we use their conjunction as the top-level output.
\end{example}

\section{Sequential composition}

In this section we show how to perform sequential composition, using the outputs of one 
protocol (producer) as input of the following one (consumer).
This is similar to the general notion of a computational pipeline.

\begin{definition}
Consider finite sets $I$, $M$, $O$ 
        (informally: global input, language in the middle, and global output, 
        the protocols working from $I$ to $M$ and from $M$ to $O$).
        Consider also finite sets $Q_1$ and $Q_2$ 
        (to be used as internal memory languages for the two protocols).
Consider two input-saving population protocols 
        $(I\times{}Q_1,Step_1,o_1)$ and $(M\times{}Q_2,Step_2,o_2)$
        (the producer and the consumer protocols, correspondingly).

Then the \emph{sequential composition} of these two protocols
is the following input-saving population protocol.
The input language is $I$. The output language is $O$. 
The internal memory language is $(Q_1\times{}M\times{}Q_2)$;
we identify $\bot$ with $(\bot,\bot,\bot)$ in the internal memory.
        The output function is $o(q_1,m,q_2) = o_2(q_2)$.

        The step relation is defined by the following procedure.
        The procedure is deterministic if the two original
        protocols 
        had functions as step relations.
        The procedure takes two states (saved input included),
        $(i^1,q_1^1,m^1,q_2^1)$ 
        and 
        $(i^2,q_1^2,m^2,q_2^2)$.
\begin{itemize}
        \item We compute the new internal memory for the first (producer) protocol 
                based on the global inputs and the old internal memory states.
                In other words, 
                we non-deterministically pick an arbitrary 
                fitting transition
                $(
                ((i^1,q_1^1),(i^2,q_1^2)),\\
                ((i^{1},q_1^{1'}),(i^{2},q_1^{2'}))
                )\in Step_1$
                if there is any, and otherwise we do nothing and take
                $(q_1^{1'},q_1^{2'})=(q_1^1,q_1^2)$.
                Note that the inputs cannot change during a step.
        \item We update the intermediate output-input as the output
                of the first (producer) protocol.
                In other words, we set
                $m^{1'}=o_1(i^{1},q_1^{1'})$
                and 
                $m^{2'}=o_1(i^{2},q_1^{2'})$.
        \item We compute the new internal memory for the second (consumer) protocol,
                based on the intermediate output-input as the input.
                In other words, 
                we non-deterministically pick an arbitrary 
                fitting transition
                $(
                ((m^{1'},q_2^1),(m^{2'},q_2^2)),\\
                ((m^{1'},q_2^{1'}),(m^{2'},q_2^{2'}))
                )\in Step_2$
                if there is any, and otherwise we 
                do nothing and take
                $(q_2^{1'},q_2^{2'})=(q_2^1,q_2^2)$.
                Note that the input cannot change but we use its updated value.
        \item The new states are 
                $(i^{1},q_1^{1'},m^{1'},q_2^{1'})$ 
        and 
                $(i^{2},q_1^{2'},m^{2'},q_2^{2'})$.
                (The input doesn't change, as required for an input-saving protocol).
\end{itemize}
\end{definition}

We also want to combine specifications.

\begin{definition}
        The composition of relations $\varphi_1\subseteq \mathcal{I}\times\mathcal{M}$
        and $\varphi_2\subseteq \mathcal{M}\times\mathcal{O}$,
        is the relation $\varphi\subseteq\mathcal{I}\times\mathcal{O} = 
        \{(I,O)\mid\exists M: \varphi_1(I,M)\wedge\varphi_2(M,O)\}$.
        This is compatible with composition of relations 
        as (multi-valued) functions.

        The composition of specifications composes the two components correspondingly.
\end{definition}

\begin{theorem}
        \label{theorem:composition}
Sequential composition of input-saving population protocols
implementing two specifications
implements the composition of the specifications as relations.
\end{theorem}

The basic idea is that things that are allowed to happen
during a fair execution with reconfigurations of the composition protocol
correspond to things that are allowed to happen during each of the two
protocols executed independently; and once things stabilise, we can 
recover the intermediate values from the intermediate output-input components 
of the internal memory.

\begin{toappendix}
\begin{proof}[of \cref{theorem:composition}]
Let the original protocols be 
        $(Q_1,Step_1)$ and $(Q_2,Step_2)$
        with 
        $Q_{1}=I\times{}Q_{1i}$
        and 
        $Q_{2}=M\times{}Q_{2i}$,
        and let the output functions be
        $o_1: Q_1\to{}M$
        and $o_2:Q_2\to{}O$.
        Then the composition $(Q,Step)$
        has $Q=Q_1\times{}Q_2$
        and the output function 
        $o:Q\to{}O$ 
        mapping $(i,q_1,m,q_2)$ to $o_2(m,q_2)$.
        Let the original specifications be $(\varphi_1,\psi_1)$ and
        $(\varphi_2,\psi_2)$ with composition $(\varphi,\psi)$.

        Consider a fair execution with reconfiguration 
        $(C_n)$ of the composition protocol $(Q,Step)$.
        Consider the projections
        $\pi_{Q_1}: (i,q_1,m,q_2)\mapsto (i,q_1)$,
        $\pi_{Q_2}: (i,q_1,m,q_2)\mapsto (m,q_2)$,
        $\pi_{IMO}: (i,q_1,m,q_2)\mapsto (i,m,o(i,q_1,m,q_2)=o_2(m,q_2))$.
        We also apply these projections to configurations, steps,
        and multisets of states in a natural way.

        Let us check that if we apply the first projection $\pi_{Q_1}$ 
        to the execution $(C_n)$,
        we obtain a fair execution with reconfiguration  $(\pi_{Q_1}(C_n))$
        of the first original protocol $(Q_1, Step_1)$.

        The reconfigurations correspond to similar reconfigurations after the projection,
        as the projection of the shutdown state $\pi_{Q_1}(\bot,\bot,\bot,\bot)=(\bot,\bot)$
        is the shutdown state,
        and the input of the projection is the same as the input of the original state.
        Any step of the composition protocol $(Q,Step)$ first performs a change that is projected
        to a step of the first original protocol $(Q_1,Step_1)$ or to
        preserving all the states, then proceeds to do
        changes to the components ignored by the projection $\pi_{Q_1}$.

        Moreover, at each moment every possible step of the first original protocol $(Q_1,Step_1)$
        can be implemented as a projection via $\pi_{Q_1}$ of 
        a possible step of the composition protocol $(Q,Step)$.
        Therefore the projection of a fair execution is also fair,
        and the projection of a fair execution with reconfiguration is a fair 
        execution with reconfiguration.

        As the projection $(\pi_{Q_1}(C_n))$ is 
        a fair execution with reconfiguration of the first
        original protocol, the outputs according to 
        the first original protocol eventually stabilise. 
        Moreover, if the input of an agent stabilises at $\bot$,
        so do its internal memory and output, as 
        the first original protocol $(Q_1,Step_1)$
        respects shutdown requests.

        Note also that after each step of the composition protocol $(Q,Step)$,
        if the state is $(i,q_1,m,q_2)$, then its third component $m$
        is the output of the first protocol for the projection 
        $m=o_1((i,q_1))$.
        As the first original protocol $(Q_1,Step_1)$ implements 
        the first specification $\varphi_1$,
        the stabilised multiset of inputs and
        the stabilised multiset of third components
        will be accepted by the relation $\varphi_1$.

        We now consider the result of applying the first projection $\pi_{Q_2}$ 
        to the execution $(C_n)$.
        The sequence $(\pi_{Q_2}(C_n))$ is not an execution with reconfiguration
        of the second original protocol $(Q_2, Step_2)$, but we are going to check
        that it is a subsequence of a fair execution with reconfiguration.
        The reconfigurations adding or removing shutdown agents do the same after 
        projection. 
        The reconfigurations changing the inputs (in terms of the composition
        protocol $Q, Step$) do not change the projection at all. 
        Now consider normal execution steps.
        The first part of a step (simulating the first protocol)
        does not change anything in terms of $(\pi_{Q_2}(C_n))$.
        The second part, updating the intermediate output-input, 
        is an input-changing reconfiguration after projection. 
        The third part of a step 
        (simulation of the second protocol)
        is projected to a normal execution step.
        Thus the entire execution step is projected 
        to a reconfiguration step followed by an execution step.

        Furthermore, we have previously shown that the third components
        (the outputs of the first protocol) stabilise.
        Therefore after some point only normal execution steps
        happen in the projection.
        As every possible step of the second protocol $(Q_2,Step_2)$ can be 
        simulated by a step of the composition protocol $(Q,Step)$,
        the suffix of the projection after the stabilisation of inputs
        is fair.
        As the second protocol $(Q_2,Step_2)$ implements a specification,
        each agent's output in $(\pi_{Q_2}(C_n))$ only changes a finite
        number of times.

        Consider the sequence of some agent's states
        $(i^t,q_1^t,m^t,q_2^t)$.

        If the eventual stable value of $i^t$ 
        is the shutdown request $\bot$, 
        then after some point $q_1^t$ will always be $\bot$
        as the first protocol $(Q_1,Step_1)$ respects shutdown requests, 
        and 
        $\pi_{Q_1}((i^t,q_1^t,m^t,q_2^t))=(i^t,q_1^t)$
        is the sequence of states of an agent in a fair
        execution with reconfiguration of the first protocol.
        From this point on, $i^t=q_1^t=\bot$ thus
        $m^t=o_1((i^t,q_1^t))=o_1((\bot,\bot))=\bot$.
        As the second protocol $(Q_2,Step_2)$
        respects shutdown requests and 
        $\pi_{Q_2}((i^t,q_1^t,m^t,q_2^t))=(m^t,q_2^t)$
        is a subsequence of the sequence of states of an agent
        in a fair execution with reconfiguration of the second protocol,
        from some point on $q_2^t=\bot$, and so the entire state 
        $(i^t,q_1^t,m^t,q_2^t)=(\bot,\bot,\bot,\bot)$
        is the shutdown state.
        Therefore the composition protocol $(Q,Step)$ respects shutdown requests.

        Otherwise, the output of the second projection
        (according to the second protocol),
        $o_2(\pi_{Q_2}((i^t,q_1^t,m^t,q_2^t)))=o_2((m^t,q_2^t))$
        stabilises as shown above, thus
        the output according to the composition protocol also stabilises
        as it is the same:
        $o((i^t,q_1^t,m^t,q_2^t))= o_2((m^t,q_2^t))$.

        Finally, consider the sequence $(\pi_{IMO}(C_n))$.
        We know that eventually it stops changing forever.
        The multisets of the first two components are accepted by $\varphi_1$,
        the multisets of the last two components are accepted by $\varphi_2$,
        thus the multisets of the first and the last components
        are accepted by the composition $\varphi$ of $\varphi_1$ and $\varphi_2$.
        Furthermore, for each agent the first two components
        satisfy $\psi_1$ while the last two component satisfy $\psi_2$
        thus the first and the last component satisfy the composition $\psi$.
        
        Hence the composition protocol $(Q,Step)$
        respects shutdown requests; has each agent's output
        change only a finite number of times
        in a fair execution with reconfiguration;
        has each agent's output eventually be compatible
        with its input with respect to the composition of $\psi_1$ and $\psi_2$;
        and has the eventual multisets of inputs 
        and outputs satisfy the composition of the relations
        $\varphi_1$ and $\varphi_2$.
        Thus it implements the composition of specifications.
\end{proof}
\end{toappendix}

\section{Expressive power}

In this section we study which specifications can be implemented by 
input-saving population protocols.
We show that these specifications are the semilinear
(Presburger-definable) ones.

We use both equivalent definitions of these sets: 
they are definable as finite unions of linear sets,
where a linear set is the set of points obtained from a base vector
by repeated addition of the vectors from a fixed finite set of periods;
and they are definable by boolean combination of 
equalities, inequalities, and fixed-modulo modular equalities of integer
linear combinations of coordinates.

Strictly speaking, a protocol implementing a specification $(\varphi,\psi)$
implements a disjunction $(\varphi\vee\varphi',\psi)$ for any relation $\varphi'$.
We will call the former specification stronger than the latter.
We will prove that for each specification implemented by
a protocol there is a stronger semilinear specification implemented
by some other protocol.

\begin{definition}
        A specification $(\varphi,\psi)$ is \emph{semilinear} 
        if $\varphi$ 
        is semilinear when considered as a predicate on 
        a tuple of multiplicities of possible input and output values.
\end{definition}

\begin{theorem}
        \label{theorem:implement-semilinear}
        Every semilinear specification 
        which can in principle be satisfied for 
        each multiset of inputs
        is implemented by some 
        input-saving population protocol.

        Here a specification $(\varphi,\psi)$ can be satisfied for a multiset $I$ of inputs
        if there exists a multiset $P$ of pairs satisfying $\psi$,
        such that the multiset of the first components of all pairs in $P$ is $I$,
        the multiset of the second components of pairs in $P$ is some multiset $O$,
        and the multisets $I$ and $O$ of the first and the second components 
        satisfy $\varphi$. 

\end{theorem}

The basic (and very inefficient) idea is that we have already seen 
protocols that are enough to compute semilinear predicates,
so we can guess the output assignment and verify that it is suitable.

\begin{toappendix}
        \begin{proof}[of \cref{theorem:implement-semilinear}]

        Note that the example protocols from \cref{example:sum}
        (or their modulo variants)
        in composition with \cref{example:presence} can reach consensus
        on the value of
        atomic predicates from the definition of Presburger sets,
        while parallel composition allows to compute the boolean combinations.

        Now we can run the following protocol: the internal memory language
                is the Cartesian product $(O\cup\{\bot\})\times{}Q_{check}$ 
                of the output language $O$ 
                demanded by the specification
                and the internal memory language $Q_{check}$
                of a protocol that verifies that a specific 
                assignment of inputs from the input language $I$
                and the outputs from the output language $O$
                satisfies the given specification $(\varphi,\psi)$.
                The intuition is that the choice of the value from $O$ 
                is the guessed eventual output.
                The output function is $o(o_{guess},q_{check})=o_{guess}$.
                As usual we identify $(\bot,\bot)$ with $\bot$.

                The transitions are non-deterministically chosen
                between doing nothing,
                doing the same as for the verification 
                (without changing the guessed outputs),
                and when both agents have output «No»
                for the verification, 
                setting the guessed outputs to arbitrary
                values from $O$; also, if there is a shutdown
                request in the input, the guessed output component of
                the internal memory is changed to $\bot$.

                For the verification protocol an execution with 
                reconfiguration is simulated (the changes
                of guessed outputs correspond to input changes).
                Once the global inputs stop changing, 
                and shutdown requests are processed,
                it is always possible to not change the guessed outputs
                long enough that the verification process converges,
                then change all the outputs to suitable values if 
                the verification rejects the previous assignment, 
                then run the verification process again without changing 
                the guesses, then observe that the guesses can no longer be
                changed and the verification has stabilised.

                Thus the protocol can always reach a configuration
                with good outputs, and in a fair execution it will do so.
        \end{proof}
\end{toappendix}

\begin{theorem}
        \label{theorem:extract-semilinear}
        Every specification implemented by 
        an input-saving population protocol
        is weaker (accepts more multisets of input-output pairs)
        than some implementable semilinear specification.
\end{theorem}

The basic idea is to consider two notions of reachability:
with and without reconfiguration. 
Due to respect of shutdown request, 
execution with reconfiguration can always return to a configuration
of size $1$.
Consider a bottom strongly connected component (BSCC)
for reachability between size-$1$ configurations.
Configurations of arbitrary size reachable from this 
BSCC
via executions with reconfiguration have mutual reachability 
with this BSCC and thus form a semilinear set 
\cite{DBLP:journals/corr/abs-1301-4874}.
BSCCs of executions without reconfiguration
form a semilinear set \cite{esparza_et_al:LIPIcs:2015:5377};
those of them that are reachable (with reconfiguration) from the chosen 
size-$1$ reconfiguration BSCC 
form a semilinear set as an intersection of two semilinear sets. 
It turns out that this set has to be allowed by the specification
and also have at least one configuration for each multiset of inputs.

\begin{toappendix}
\begin{proof}[of \cref{theorem:extract-semilinear}]
        Consider a protocol $(Q, Step)$ with input language $I$ and output language $O$,
        the output function being $o:Q\to{}O$.

        Consider the set $\mathcal{C}^{\bot}$ of all configurations with $1$ agent
        and its input $\bot$, and the relation $R^r$ for reachability between such configurations
        via executions with reconfiguration. Consider any bottom strongly connected component
        $\mathcal{B}^\bot$
        of the relation $R^r$ on $\mathcal{C}^\bot$ that is reachable from $(\bot,\bot)$,
        and some configuration in it, $B\in\mathcal{B}^\bot$.
        Consider also the relation $R^0$ denoting reachability without reconfiguration
        (on all configurations).
        Let $\mathcal{S}$ be the union of all of its bottom strongly connected components
        (BSCCs). It is a semilinear set \cite{esparza_et_al:LIPIcs:2015:5377}.

        Every configuration reachable from $B$ is reachable from empty configuration
        via an execution with reconfiguration, as $(\bot,\bot)$ can be obtained by
        an agent-adding reconfiguration, and by assumption $\mathcal{B}^\bot$
        and in particular $B$ is reachable from $(\bot,\bot)$.
        Moreover, if a configuration $C$ is reachable from $B$,
        reconfiguration replacing every input with a shutdown request $\bot$
        followed by a fair execution without reconfiguration until all the agents
        (except at most one shut down) and agent-removal reconfigurations
        allows to return to some configuration $B'$ in $\mathcal{C}^\bot$, i.e. a configuration
        with one agent and its input $\bot$. Note that $B'$ is reachable (with reconfiguration)
        from $B$ thus it has to be in the BSCC $\mathcal{B}$, and then $B$ is reachable from
        $B'$. By transitivity, $B$ and $C$ are mutually reachable: 
        $B \, R^r \, C \, R^r \, B' \, R^r \, B$.
        Then the set $\mathcal{C}$ of all configurations reachable from $B$
        is the set of all configurations mutually reachable with $B$ via
        executions with reconfiguration.
        As both the steps and the reconfigurations can be modelled 
        in the standard way via vector addition systems, we can apply 
        the result \cite{DBLP:journals/corr/abs-1301-4874}
        that $\mathcal{C}$ is a semilinear set.

        Consider now any configuration $C^S$ from the intersection
        $\mathcal{C}\cap\mathcal{S}$ of the configurations reachable from $B$
        with reconfigurations, and the configurations in BSCCs with respect to
        executions without reconfigurations.
        There is a fair execution with reconfiguration that reaches 
        $C^S$ infinitely often (by reaching it once and then never
        doing any further reconfiguration).
        Thus $C^S$ has to satisfy the specification $(\varphi,\psi)$.

        Now observe that $\mathcal{C}\cap\mathcal{S}$ is a semilinear set
        as an intersection of two semilinear sets.
        It has to contain configurations with all possible multisets of inputs,
        as we can obtain the necessary multiset of inputs 
        via reconfiguration from $B$,
        and then a fair execution without reconfiguration
        would not change the inputs but would need to reach 
        some BSCC of $R^0$\cite{esparza_et_al:LIPIcs:2015:5377}.
        That BSCC would be a part of $\mathcal{S}$, and the 
        overall execution with reconfiguration would certify 
        that it would also be in $\mathcal{C}$.

        Thus we have proven that 
$\mathcal{C}\cap\mathcal{S}$ is a semilinear set 
        of some of the configurations
satisfying the specification $(\varphi,\psi)$ 
with at least one configuration for every possible multiset of inputs.
The relation between multisets of inputs and outputs
observed in $\mathcal{C}\cap\mathcal{S}$
is an implementable semilinear specification stronger 
than (or equivalent to) $(\varphi,\psi)$, as required.
\end{proof}
\end{toappendix}

\section{Conclusion and future directions}

We have presented an extension of the population protocols model
with a better support for modular design and synthesis of the protocols,
as well as for a wider range of applications.

We have also established that the specifications that 
can be implemented this way are essentially the semilinear ones.

However, the constructions used are quite inefficient; 
it might be of interest to adapt to our model
(and possibly extend beyond computing predicates) the existing 
fast and succinct protocols \cite{DBLP:journals/jcss/CzernerGHE24}.
However, even a single-exponential-state-count polynomial-convergence-time
construction for an arbitrary semilinear specification would be of interest.

\begin{credits}
\subsubsection{\ackname} 
I would like to thank Javier Esparza, Roland Guttenberg, Jérôme Leroux and Chana Weil-Kennedy for useful discussions.
I am also grateful to the anonymous reviewers of this and previous versions 
for their valuable feedback and advice on presentation.

The project has been supported by a European Research Council (ERC) project under the European Union’s Horizon 2020 research and innovation programme grant agreement No 787367 (PaVeS)
This work has been supported by France National Research Agency (ANR) grant ANR-23-CE48-0005 (PaVeDyS).

\subsubsection{\discintname}
The author(s) have no competing interests to declare that are relevant to the content of this article. 
\end{credits}

\bibliography{construction-population-protocols.bib}
\appendix
\end{document}